\documentclass[12pt]{article}

\usepackage{amsfonts}
\usepackage[all]{xy}
\usepackage{amsmath}
\usepackage{amsthm}
\usepackage{amssymb}

 \theoremstyle{plain}
\newtheorem{thm}{Theorem}[section]

\newtheorem{proposition}[thm]{Proposition}
\theoremstyle{definition}

\theoremstyle{remarks}
\newtheorem{remarks}[thm]{Remarks}

\title{The Spin-Statistics Relation in Nonrelativistic
Quantum Mechanics and Projective Modules}
\author{N.A. Papadopoulos\footnote{Johannes-Gutenberg-Universit\"at, Institut f\"ur
Physik-ThEP. Staudinger Weg 7 Mainz, D-55128, Germany.} , M.
Paschke\footnote{Max Planck Institute for Mathematics in the
Sciences. Inselstrasse 22-26 Leipzig, D-04103, Germany }, A.F.
Reyes$^*$
 and F. Scheck$^*$}

\date{}
\begin{document}
\maketitle
\newcommand{\A}{{\mathcal{A}}}
\newcommand{\BbP}{{\mathbb P}}
\newcommand{\BbR}{{\mathbb R}}
\newcommand{\BbC}{{\mathbb C}}
\newcommand{\BbZ}{{\mathbb Z}}
\newcommand{\BbN}{{\mathbb N}}

\begin{abstract}
In this work we consider non-relativistic quantum mechanics, obtained
from a classical configuration space $\mathcal Q$ of
indistinguishable particles.
Following an approach proposed in \cite{P},
wave functions are regarded as elements of suitable projective modules over
$C(\mathcal{Q})$.
We  take furthermore into account the $G$-Theory point of
view (cf. \cite{HPRS,S}) where the role of group action
is particularly emphasized.
 As an example illustrating the method,  the case of two
particles is worked out in detail.
Previous works (cf.\cite{BR1,BR2})  aiming at a proof of a spin-statistics
theorem for non-relativistic quantum mechanics are re-considered from the
point of view of our approach, enabling us to clarify several points.
\end{abstract}
\newpage
\section{Introduction}
Being a  consequence of  the general principles of relativistic QFT,
 the  spin-statistics (SS) theorem   has found rigorous proofs in the context
of axiomatic QFT, as well as in that of  algebraic  QFT.

In spite of various efforts, this has not been  the case  in non relativistic
quantum mechanics (NRQM).
However, it would be interesting to find such a proof, which does not rely as
heavily on concepts of relativistic QFT as the established ones, for several
reasons. There are many examples
 of  phenomena
taking place outside the relativistic realm which depend essentially on the
SS relation for its description, making such a proof desirable. On the other
hand, this new  sought-after approach to SS could
also be of benefit for the understanding of QFT itself. For instance, a proof
which does not make use of the complexification of the  full
Lorentz group could provide hints towards the understanding of SS in more
general situations, such as theories where a background gravitational
field is present. It could also provide guidance for the development of
theories on non commutative spaces.
There is  also a motivation coming from the idea that
{\it quantum indistinguishability}, if correctly incorporated into quantum theory,
might lead to a better understanding of SS.

Usually, arguments along these
lines are based on what can be called the {\it  ``configuration space approach''}, since
in one form or another they make use of configuration space
techniques.
For example, in one of the first works of this kind,
 Laidlaw and DeWitt  found out
in \cite{LDeW} that when applying the path
integral formalism to a system consisting  of
a finite number of  non-relativistic, identical spin
zero particles in three spatial dimensions, the topology of the
corresponding configuration space imposed certain restrictions on the
 propagator. From this, they were able to deduce that only particles
obeying Fermi or Bose statistics were allowed (note that this Fermi-Bose
alternative is an input in the standard proofs of axiomatic QFT).
Leinaas and Myrheim   considered a similar situation
  in \cite{LM}, in an analysis
that was motivated by the relevance of indistinguishability
to Gibb's paradox. They argued that if a quantum theory is
obtained after a process of quantization on a classical configuration space
$\mathcal Q$,
then  indistinguishability should be incorporated
in $\mathcal Q$
right from the beginning.
Mathematically, they considered wave functions to be
sections of  vector bundles on a  space
where permuted configurations were identified. By physical reasons,
these vector  bundles should be equipped with a flat connection,
 whose holonomy was shown to describe the effect of particle exchange.
They reproduced the results of \cite{LDeW} by  obtaining the Fermi-Bose
alternative in three dimensional space for spinless particles.
 But, in  addition, they also
 found that in one and two dimensions the statistics parameter
could, in principle,  take infinitely many values.
In that same work, they remarked that their results could provide a
geometrical basis for the derivation of the SS theorem.
A lot of work based on this kind of ``topological arguments'' has
been done since then, but most of the results that have been obtained
are based on assumptions which go beyond NRQM.
It should also be said that  in several cases the argumentations remain
at a classical level, having no clear interpretation in terms of
quantum mechanics.

More recently, Berry and Robbins provided in \cite{BR1} an explicit
construction in which the quantum mechanics of two identical particles is
formulated along the lines described in \cite{LM}, leading  to  the physically
correct SS relation. Since their result is based on a particular construction
of what they call a ``transported spin basis'', one cannot consider it as a
proof from first principles -as the authors themselves have recognized, since
there are various alternatives for the construction of such a spin basis
leading to different statistics signs (cf.\cite{BR2})-. Nevertheless, the
construction is interesting for its own sake, and raises several questions that
deserve to be considered.

In this paper we consider the SS relation (for two particles) from an
algebraic point of view, by studying projective modules over
$C(\mathcal{Q})$ (by the Serre-Swan theorem these modules can be interpreted as
modules of sections on vector bundles over $\mathcal{Q}$). Additionally, we
assume  the $G$-Theory point of view  for the consideration of the symmetries
of the problem. This allows us to arrive at a  precise and explicit formulation
of the SS problem, in which various known  results can be reproduced in a clear
and efficient way. As an example illustrating the relevance of the proposed
approach, we make a comparison with the  Berry-Robbins construction and show
how various points  can be clarified.

The paper is organized as follows.   In section {\bf 2} we
present the method through a detailed discussion of the two particle case.
In section {\bf 3} the work of Berry and Robbins is briefly reviewed
and a comparison with the approach presented in section {\bf 2} is made.
Finally, we present in section {\bf 4} a brief discussion
of the results.

\section{Projective modules and the configuration space approach
to SS: an example.}

The classical configuration space of a system of $N$ identical
particles moving in $\BbR ^3$ is defined as a quotient space,
$\mathcal{Q}_N=\mathcal{\widetilde Q}_N/S_N$, obtained from the natural action
of the permutation group $S_N$ on  the space
\begin{equation}\label{eq:Q_N}
\mathcal{\widetilde Q}_N=\{(r_1,\ldots,r_N)\in \BbR^{3N}|r_i\ne r_j
\hbox{ for all pairs} (i,j) \}.
\end{equation}
The non-coincidence condition  $r_i\ne r_j$ is included  in the definition of
$\mathcal{\widetilde Q}_N$ in order to make $\mathcal{Q}_N$ a manifold.
Following \cite{LM}, we consider wave functions to be given by square
integrable sections of some vector bundle on $\mathcal{Q}_N$.

 In this work we will restrict ourselves
to the case $N=2$. Here,  after performing a transformation to center
of mass and
relative coordinates,  one sees that $\mathcal{Q }_2$ is
of the same homotopy type as a two-sphere $S^2$, this latter representing the space
of normalized relative coordinates of the two particles. Under exchange,
the relative coordinate $r$ goes to $-r$, so that after quotienting out
by the action of $S_2\cong\BbZ_2$, we obtain a projective plane.
For our purpose it is therefore enough to consider
$\mathcal{\widetilde Q}_2= S^2$ and
 $\mathcal{Q}_2=\BbR\BbP^2$ for the configuration space.
The sphere will be  considered as embedded in $\BbR^3$. Points on it
will be denoted by $x=(x_1,x_2,x_3)$ and, accordingly,
points in the projective plane will be denoted  by
$\left[ x\right]=\{x,-x\}$.

Define now $\mathcal{A}:=C(S^2)$. The $\BbZ_2$-action on $S^2$ induces one
on  $\mathcal{A}$, leading to a decomposition into subspaces
of even and odd functions:
\begin{equation}\label{A=A++A-}
\A=\A_{+}\oplus\A_{-}.
\end{equation}
 $\A_{+}$, the subalgebra of even functions, is easily seen to be
isomorphic to $C(\BbR\BbP^2)$. For the description of the spin degrees of
freedom, we will certainly need a representation of $SU(2)$ taking into account
the transformation properties of the  wave functions under rotations. Since
$S^2$ is a homogeneous space for $SU(2)$ and consequently $\mathcal A$ carries
an $SU(2)$ representation, the isomorphism $\A_{+}\cong C(\BbR\BbP^2)$ offers
the possibility of constructing projective modules corresponding to
$SU(2)$-equivariant bundles over $\BbR\BbP^2$, just by exploiting the
rotational symmetry of the sphere. The construction of a projective module
having these properties has already been carried out in \cite{P}, but for the
sake of completeness we reproduce it here. Let us denote with  $V^j$, as usual,
the $(2j+1)$-dimensional irreducible $SU(2)$ representation, so that $\A\cong
\bigoplus_{j\in \BbN_0}V^j$. Consider  the tensor product representation
$\A\otimes V^1$. Decomposing it into irreducibles, we obtain:
\[
\A\otimes V^1 \cong \left(\bigoplus_{j\in N_0}V^j\right)\otimes V^1
\cong V^1\oplus(V^0\oplus V^1\oplus V^2 )\oplus (V^1\oplus V^2\oplus
V^3)\oplus\cdots
\]
Note that the trivial representation $V^0$ appears only once
in the decomposition.
Thus,  there is a unique scalar element, up to normalization,
with respect to this representation.
Using $\A\otimes V^1\cong\A^3$, this scalar element is given, in terms
of spherical harmonics,  by the (normalized) vector
\begin{equation}\label{eq:|Psi>}
|\psi\rangle:=\sqrt{\frac{4\pi}{3}} \left(\begin{array}{c}
    Y_{1,1} \\
    -Y_{1,0}\\
    Y_{1,-1}
  \end{array}\right)
\end{equation}
This $\A$-valued vector has remarkable properties.
For instance, we have the following result.
\begin{proposition}\label{prop:1}(cf.\cite{P}) Define a projector on $\A^3$ by
$ p:=|\psi\rangle \langle\psi|. $ Then, the following isomorphism of
$\A_{+}$-modules holds: $p(\A_{+}^3)\cong \A_{-}$.
\end{proposition}
By the Serre-Swan theorem, there is a line bundle $L_-$ on $\BbR\BbP^2$
whose module of sections is isomorphic to $p(\A_+^3)$.
Since  $p(\A_+^3)$ is, by construction, $SU(2)$-equivariant (cf.\cite{P}),
it follows that  $L_-$ is also $SU(2)$-equivariant.
In the following proposition we give a proof of this fact in a form
suitable for the applications in the next section.
\begin{proposition}\label{prop:2}
The line bundle $L_-$ is $SU(2)$-equivariant.
\end{proposition}
\begin{proof}
 First, note  that  the total space $E(L_-)$ of the bundle is defined,
as a {\it set}, as
$E(L_-):=\{(\left[ x\right],\lambda |\psi(x)\rangle)\in \BbR P^2\times\BbC^3\,\;
|\;\; \lambda\in \BbC, x \in \left[ x \right]\,\}.$
Therefore,   $L_-$ can be regarded as a subbundle of the trivial bundle
 $\BbR\BbP^2\times \BbC^3$, with projection map
$\pi(\left[ x \right], \lambda |\psi(x)\rangle):=\left[ x \right]$.
Since  $|\psi\rangle$ is invariant under
$SU(2)$ we know  that for every $g\in SU(2)$ and $x\in S^2$ the relation
${\mathcal D}^{(1)}(g)|\psi(x)\rangle = |\psi(g\cdot x)\rangle$
holds. This can be used to define an $SU(2)$ action on $E(L_-)$. Indeed,
since $|\psi(x)\rangle\,(\,=-|\psi(-x)\rangle\,)$ spans the fiber over
$\left[x \right]$, we see that the action of an element $g\in SU(2)$
on
$y=(\left[ x\right],\lambda |\psi(x)\rangle)\in \pi^{-1}(\left[ x \right])$
can be correctly defined by setting
$\tau_g (y):=(\left[g x \right],\lambda | \psi(g\cdot x)\rangle )$.
\end{proof}

\begin{remarks}
$\,$\\
\begin{enumerate}

\item It is well known that -up to equivalence- there are only two
complex line bundles on $\BbR\BbP^2$. One of them is the trivial one, $L_+$,
and the other is $L_-$. The corresponding modules of sections
$\Gamma(L_\pm)$ are isomorphic, respectively, to $\A_\pm$. Higher rank bundles
can always be written as sums of these bundles.
\item  The Grassmann connection naturally associated to $p(\A^3_+)$
 is defined as
$\nabla=p\,dp$. It can be shown, by direct computation, that $\nabla$ has
vanishing curvature. The corresponding holonomy group is  $\BbZ_2$.
\item
The choice of $\A_-$ as  Hilbert space corresponds, according to the discussion
in the introduction,  to the description of a system of two {\it identical}
spin zero particles obeying Fermi statistics. Those obeying Bose statistics are
of course described by $\A_+$ and, by the first remark above, there are no
other possibilities. Note that this result is obtained from  an intrinsic
treatment of indistinguishability, where no use of a symmetrization postulate
is made.
\end{enumerate}
\end{remarks}
We thus see how the  Fermi-Bose alternative for scalar particles is obtained as
a direct consequence of the topology of the configuration space. This is
 a well known result, but we have discussed it in order to illustrate the
method. The usefulness of our approach will be clear when we consider  higher
values of the spin, since in that case the Fermi-Bose alternative has no direct
relation to the topology of the configuration space: the $SU(2)$-equivariance
of bundles on it must be used. In fact, in the general case the Fermi-Bose
alternative follows from the requirement of a well defined transformation law
for the wave functions under rotations,
 compatible with the exchange of particles (cf.\cite{PRS,R}).

Now we consider the  relation between the $SU(2)$ action on $L_-$
and  parallel transport with respect to $\nabla=p\,dp$.
For the proof of the following proposition, it is convenient to
consider the equivalent projector $\tilde p:=U^\dagger p\, U$,
where
\[
U=\left( \begin{array}{ccc}
            1/\sqrt{2} & -i/\sqrt{2} &  0 \\
               0       &    0        & -1 \\
           -1/\sqrt{2} & -i/\sqrt{2} &  0 \\
           \end{array}
  \right).
\]
All the previous formulae for $p$ (or $L_-$) in terms of
$|\psi\rangle$ remain valid for   $\tilde p$ upon replacing
$|\psi\rangle$ by $|\phi\rangle:=U^\dagger |\psi\rangle$ in them.
\begin{proposition}\label{prop:3}
Let
$\gamma: \left[0,1\right]\rightarrow SU(2)$ be a smooth path such that
$\gamma(0)=e$. Given
$\left[x^{(0)}\right]\equiv \{x^{(0)}, - x^{(0)} \}\in \BbR\BbP^2$ and
$y\in E(L_-)$ with  $\pi(y)=\left[x^{(0)}\right]$, define
$\alpha:\left[0,1\right]\rightarrow \BbR\BbP^2$ through
$\alpha(t)=\left[ \gamma(t)\cdot x^{(0)}\right]$. Then
it follows that
$t\mapsto\tau_{\gamma(t)}(y)$ is a section along
$\alpha$, parallel with respect to the
connection $\tilde\nabla=\tilde p\,d\tilde p$.
\end{proposition}
\begin{proof}
Pick one representative of  $\left[x^{(0)}\right]$, say $x^{(0)}$. From the
description of $E(L_-)$ given in the proof of proposition \ref{prop:2},
 we know that
there exists  $\lambda\in \BbC$ such that $y=(\,\left[x^{(0)}\right], \lambda
|\phi(x^{(0)})\rangle\,)$ (this $\lambda$ is unique, since a choice of
representative has been made). By letting $SU(2)$ act on the sphere, we obtain
three real functions $t\mapsto x_i(t)$, $i\in\{1,2,3\}$, namely, the components
of $x(t):=\gamma(t)\cdot x^{(0)}$. Now, note that the columns of the projector
$\tilde p$ give place to sections $e_i\in \Gamma(L_-)$, $i\in\{1,2,3\}$,  that
are generators for  the module. Writing $e_i(t)$ for $e_i \circ \alpha (t)$, we
can define $s(t):=\sum_{i=1}^3 x_i(t)e_i(t)$,
  a section along $\alpha$.
It is easily verified that $s(t)\equiv\tau_{\gamma(t)}(y)$. That $s$
is parallel along $\alpha$ follows directly from the explicit form
of the connection: $\tilde \nabla e_i = \sum_{j=1}^3 da_{ij} \otimes
e_j$, where $a_{ij}(\left[x\right]):=x_ix_j$.
\end{proof}
We finish this section with some comments about the $N>2$
case.
Recall that because of proposition \ref{prop:1} we may regard functions on
$\widetilde{\mathcal Q}_2$
as sections on (flat) vector bundles over $\mathcal Q_2$.
Looking back at equation (\ref{A=A++A-}), we realize that
the decomposition of $C(\widetilde{\mathcal Q}_2)$ into
$C({\mathcal Q}_2)$-submodules induced by the $\BbZ_2$
action on it provides a complete description
of all flat bundles over ${\mathcal Q}_2$.
This assertion remains valid for $N>2$:
\begin{proposition}\label{prop:N} The
$C({\mathcal Q}_N)$-submodules of the algebra
 $C(\widetilde{\mathcal Q}_N)$ obtained from
the $S_N$ action on  $\widetilde{\mathcal Q}_N$
are finitely generated and projective and
the natural connections associated to them are flat.
\end{proposition}
Since only the case $N$=2 will be needed for the discussion
of the next sections,
the reader is referred elsewhere \cite{R} for the proof of the  proposition.
A remark on how the construction of proposition  \ref{prop:1}
may be extended
to arbitrary $N$ is however in order, because
the way we have decomposed $\mathcal A$ into $\mathcal A_+$-projective
modules (the case $N$=2),
depends strongly on the natural $SU(2)$-action available in this case.
For general $N$, we have a free action of $S_N$ on
 $\widetilde{\mathcal Q}_N$ (see Eq.\ref{eq:Q_N}),
which gives place to a representation of $S_N$ on
$C(\widetilde{\mathcal Q}_N)$ (this representation
turns out to be closely related  to the
regular representation). Now, since $\widetilde{\mathcal Q}_N$
is the universal covering space of ${\mathcal Q}_N$, it is
not difficult (by means of a suitably chosen
 partition of unity) to find explicit  expressions
for the transformation properties of functions  in each irreducible
subspace
 of $C(\widetilde{\mathcal Q}_N)$.
This in turn  allows one to show that each such subspace
is, in fact, a finitely generated and projective module
over $C({\mathcal Q}_N)$.

With this result at hand, the study of the general case
and of its relevance to the SS problem can be carried
out systematically. Nevertheless, in order to obtain
a {\it proof} of the SS theorem within the present approach,
further assumptions, motivated from physics, are needed.
A proposal in this direction is being carried out
and will be published shortly \cite{PRS}.
\section{Comparison with the Berry-Robbins\\ approach.}
%
Now we proceed to make a comparison between the approach explained in the last
section and that of Berry and Robbins (BR). We will be mainly concerned with
the relation between the projector $p$ of proposition \ref{prop:1}
and the transported spin basis of BR, on one hand, and with the
singlevaluedness
assumption of BR, on the other.
Some remarks will be made about the spin operators defined
in BR in relation to proposition \ref{prop:3}.
\subsection{Relation between $p$ and the BR construction.}
We begin by briefly recalling the essential points of the BR construction.
We refer the reader to their work (cf.\cite{BR1,BR2})
for details on the construction
 and also  for the notation, which we  follow here.
Consider two identical particles of spin $S$. Let the label $M$ stand for the
-ordered- pair of  eigenvalues $\{m_1,m_2\}$ of the spin of the two particles
in a given direction. If $M$ corresponds to a given pair $\{m_1,m_2\}$, then
(following BR)  we denote with  $\overline{M}$ the label corresponding to the
permuted pair $\{m_2,m_1\}$.
In the BR approach, the usual spin basis $\{|M\rangle\}_M$ is replaced by a
new, position dependent one, $\{|M(r)\rangle\}_M$. The transported spin vectors
are obtained from the usual ``fixed'' ones by means of a position dependent
unitary transformation $U$, constructed with the help of Schwinger's
representation of spin, and acting on $\BbC^{N_S}$, where
$N_S=\frac{1}{6}(4S+1)(4S+2)(4S+3)$.
The main properties of the resulting basis are the following:
\begin{itemize}
\item[(i)]The map \vspace{-0.8cm}
\begin{eqnarray*}
S^2 & \longrightarrow & \BbC^{N_S}\\
r   & \longmapsto     & |M(r)\rangle:=U(r)|M\rangle
\end{eqnarray*}
is well defined and smooth for all $M$. (Note that $U$ is really an $SU(2)$
representation, so its matrix components are not functions on $S^2$. Only when
acting on the ``physical'' vectors of the form $|M\rangle$ on  $\BbC^{N_S}$,
does one obtain a vector (at each $r$) whose components can be regarded as
functions on $S^2$).
\item[(ii)] The following ``exchange'' property holds:
\begin{equation}\label{eq:exchange}
|\overline M(-r)\rangle=(-1)^{2S}|M(r)\rangle.
\end{equation}
\item[(iii)] The ``parallel transport'' condition
$\;\langle M'(r(t))|\frac{d}{dt}M(r(t))\rangle =0\;$
 is satisfied for all $M,M'$ and any smooth curve $\;t\mapsto r(t)$.
\end{itemize}
The wave function  is then  expressed in terms of the
 transported basis,
\begin{equation}\label{eq:f-onda}
 |\Psi(r)\rangle =\sum_M \Psi_M(r)|M(r)\rangle,
\end{equation}
and the following condition is {\it imposed} on it:
\begin{equation}\label{eq:singlevalued}
|\Psi(r)\rangle \stackrel{!}= |\Psi(-r)\rangle.
\end{equation}
An immediate consequence of this is that the coefficient functions must satisfy
the relation $\Psi_{\overline M}(-r)=(-1)^{2S} \Psi_M(r)$, which is the usual form
of the SS relation. The task of (\ref{eq:singlevalued}) is to incorporate the
indistinguishability of the particles in the formalism, but we
shall postpone the discussion of this point to
section \ref{sec:3.2}. Our concern for the moment is to find out ``where does the
wave function live'' because, in the words of BR, what we are doing with this
construction is
``{\it setting up quantum mechanics on a `two-spin bundle',
whose six-dimensional base is the configuration space $r_1, r_2$ with exchanged
configurations identified and coincidences $r_1=r_2$ excluded (...). The fibres
are the two-spin Hilbert spaces spanned by the transported basis
$|M(r)\rangle$. The full Hilbert space consists of global sections of the
bundle, i.e. singlevalued wave functions}''\cite{BR1}.
Therefore, the wave function
$|\Psi(r)\rangle$ should be regarded as a section of some vector bundle over
$\BbR\BbP^2$. What we want to do first is to find out which bundle
this is.

In order to accomplish this task, we perform a change from the basis
$\{|M\rangle\}_M$ to a basis of total angular momentum
$\{|J,m_J\rangle\}_{J,m_J}$,
according to the Clebsch-Gordan decomposition, and write the transported
spin basis in terms of this new basis. The bundle these
new transported vectors generate
can then be easily  identified. Let us consider, in order to be concrete,
the
$S=1/2$ case, for which $N_S=10$. A basis for the space acted on by
$U(r)$ can be written down  in terms of the four
oscillator operators (cf.\cite{BR1,BR2}) $a^\dagger_i,b^\dagger_j$ ($i,j=1,2$). We
shall use the following one:
\[
\begin{array}{ccc}
\vspace{0.2cm}
|1,1\rangle^{(-1)}=\frac{(a_1^\dagger)^2}{\sqrt{2}}|0\rangle,\hspace{0.5cm} &
|1,1\rangle^{(0)}=a_1^\dagger a_2^\dagger |0\rangle,
  & \hspace{0.5cm}
|1,1\rangle^{(1)}=\frac{(a_2^\dagger)^2}{\sqrt{2}}|0\rangle, \\
\vspace{0.2cm}
|1,0\rangle^{(-1)}=\frac{(b_1^\dagger)^2}{\sqrt{2}}|0\rangle,\hspace{0.5cm} &
|1,0\rangle^{(0)}=b_1^\dagger b_2^\dagger |0\rangle,& \hspace{0.5cm}
|1,0\rangle^{(1)}=\frac{(b_2^\dagger)^2}{\sqrt{2}}|0\rangle, \\
\vspace{0.2cm} |1,-1\rangle^{(-1)}=a_1^\dagger b_1^\dagger
|0\rangle,\hspace{0.5cm} &
 |1,-1\rangle^{(0)}=\frac{a_1^\dagger b_2^\dagger+b_1^\dagger a_2^\dagger}{\sqrt{2}}
 |0\rangle,
  &  \hspace{0.5cm}
  |1,-1\rangle^{(1)}=a_2^\dagger b_2^\dagger |0\rangle, \\
     &      |0,0\rangle=\frac{a_1^\dagger b_2^\dagger-b_1^\dagger
a_2^\dagger}{\sqrt{2}} |0\rangle.
  & \\
\end{array}
\]
For the transported spin basis one then obtains
\begin{eqnarray*}
|J=1,\, m_J\;(r)\rangle & := & U( r)|J=1, m_J\rangle^{(0)}\\
                & = &  \sum_{\mu=-1}^1 W(r)_{0,\mu}|J=1, m_J\rangle^{(\mu)},
\hspace{0.5cm}(m_J=1,0,-1),\\
\mbox{and}\hspace{3.35cm}& &\\
|J=0,\,\, 0\,\;(r)\rangle & := & U( r)|J=0,\,\, 0\,  \rangle =
                                         |J=0,\,\, 0\, \rangle,
\end{eqnarray*}
where
\[
W(\vec r):= \left(\begin{array}{ccc} \cos^2\frac{\theta}{2} &
e^{i\varphi}\frac{\sin\theta}{\sqrt{2}}
                                       &   e^{2i\varphi}\frac{\sin^2\theta}{2} \\
-e^{-i\varphi}\frac{\sin\theta}{\sqrt{2}} & \cos\theta
                                       &  e^{i\varphi}\frac{\sin\theta}{\sqrt{2}} \\
e^{-2i\varphi}\frac{\sin^2\theta}{2} &
-e^{-i\varphi}\frac{\sin\theta}{\sqrt{2}}
                                       & \cos^2\frac{\theta}{2}
\end{array}
\right).
\]
To identify the corresponding bundle, we follow the remark -quoted above- that
the transported vectors, evaluated at the point $\pm r$, span the fiber
over $\left[r\right]$. From the last equations it is clear that
the singlet component of the wave function will lie in a trivial line bundle
and that the (line) bundles corresponding to the triplet components are all
equivalent. The projection operator onto the vector space spanned
by $|J=1,\, m_J\;(r)\rangle$ will have the same form for all $m$, so we may
just define
$ P^{(J=1)}(r):=W^t( r) P_0 W^*( r)$,
 with $(P_0)_{ij}=\delta_{2,i}\delta_{2,j}$ ($i,j=1,2,3$).
This leads to:
\[
P^{(J=1)}(r)=\left(
\begin{array}{ccc}
\frac{1}{2}\sin^2\theta & -\frac{1}{\sqrt{2}}\sin\theta\cos\theta e^{-i\varphi}
                        & -\frac{1}{2}\sin^2\theta e^{-2i\varphi}\\
 -\frac{1}{\sqrt{2}}\sin\theta\cos\theta e^{i\varphi} & \cos^2\theta
                        & \frac{1}{\sqrt{2}}\sin\theta\cos\theta e^{-i\varphi}\\
 -\frac{1}{2}\sin^2\theta e^{2i\varphi} &
                       \frac{1}{\sqrt{2}}\sin\theta\cos\theta e^{i\varphi}
                        & \frac{1}{2}\sin^2\theta
\end{array}
\right).
\]
From equation (\ref{eq:|Psi>}) we see that this projector
is exactly the same as the one  defined in proposition
\ref{prop:1}: $P^{(J=1)}\equiv p$. This means that the vector
bundle in question is $L_-\oplus L_-\oplus L_-\oplus L_+$.
The wave function must therefore be a section of this bundle or,
equivalently, an element of the $\A_+$-module
$\A_-^3\oplus \A_+$. The case of
general spin can be handled in a similar way,
so we will not consider it here.
%
\subsection{The singlevaluedness condition.}\label{sec:3.2}
%
At first glance, as implied by (\ref{eq:f-onda}), the wave function is
given by a map  $S^2\rightarrow \BbC^{N_S}$. It is only because
of  (\ref{eq:singlevalued}) that we may consider its domain to be
$\BbR\BbP^2$. But, in which sense and to what extent does
the imposition of this condition really define $|\Psi(r)\rangle$ as a
section of a bundle over $\BbR\BbP^2$?

Denote with  $L^J_{m_J}$ the bundle over $S^2$ whose fiber over $r$ is the
complex line spanned by $|J,m_J(r)\rangle$ (note that, since
$|J,m_J(r)\rangle\ne 0$ for all $r$, $L^J_{m_J}$ is trivial). Put
$\Psi(r):=(r,|\Psi (r) \rangle)$ and define $\eta^S:= \oplus_{J=0}^{S}\left(
\oplus_{m_J=-J}^J L^J_{m_J}\right). $ Because of (\ref{eq:f-onda}), we have
$\Psi\in\Gamma( \eta^S )$. But $\eta^S$ is a bundle over $S^2$, not over
$\BbR\BbP^2$, as we need. A possible way out of this problem is to specify a
$\BbZ_2$-action on $\eta^S$ and then to construct the quotient   $\eta^S /
\BbZ_2$. This is justified by the following well-known fact ($M$ is a manifold
with a free $G$-action and $G$, for our purposes, a finite group):
\begin{proposition}(cf.\cite{A})\label{prop:4}
If $M$ is $G$-free $G$-vector bundles over $M$ correspond
bijectively to vector bundles over $M/G$ by
$\eta\rightarrow \eta/G$.
\end{proposition}
In the case of the {\it line} bundle $L^J_{m_J}$, there are exactly two
such possible $\BbZ_2$-structures given, say, by actions $\tau_{\pm}$.
Quotienting out by $\tau_{\pm}$ one obtains
$L^J_{m_J}/\tau_\pm \cong  L_\pm$, but there is no
{\it a priori} way of choosing between  $\tau_-$ and $\tau_+$.
Nevertheless, there is something particular in the
way $L^J_{m_J}$ has been constructed:
because of the ``exchange'' property (\ref{eq:exchange}), we have
$|J,m_J(-r)\rangle=(-1)^J|J,m_J(r)\rangle$.
This relation  suggests somehow a choice of action, according
to whether $J$ is even or odd
\footnote{This relation is  also the reason why
 the projector
$P^{(J=1)}$ defines a module over $\A_+$, since from it
we get $P^{(J=1)}(-r)=P^{(J=1)}(r)$.}.
This is in fact true, in a certain
sense (to be explained), that involves the ``singlevaluedness'' condition
(\ref{eq:singlevalued}).
But before that we have to answer the following question:
if we can pass from $\eta^S$ to a bundle over $\BbR\BbP^2$ by specifying
a $\BbZ_2$-action  on $\eta^S$ and taking the quotient, what is the
procedure to follow with $\Psi$? The answer
is easily obtained through of a reformulation of proposition
\ref{prop:4}, as explained below.

Consider  again the situation of proposition \ref{prop:4}.  Let
$q:M \rightarrow M/G$ be the quotient map.
The proposition  says that if $\eta$ is a $G$-vector bundle over $M$
(with action $\tau$), then
$\eta/\tau$ is a bundle over $M/G$ and $q^*(\eta/\tau)\cong_G\eta$. The
equivalence is an isomorphism of $G$-bundles: $\eta$ as $G$-bundle
 with respect to $\tau$, and $q^*(\eta/\tau)$ with respect to the
$G$-action naturally inherited from the pull-back operation.
On the other hand, if $\xi$ is a bundle over $M/G$, the induced
$G$-action on $q^*(\xi)$ makes it a $G$-bundle, and then
$q^*(\xi)/G\cong \xi$.
These isomorphisms  allow one to  work on  $M$,  considering
$G$-bundles on it, in order to describe bundles on $M/G$.
 But of course  we must always take the additional structure carried
by $\eta$ into account, if we want to ``regard'' it as a bundle over $M/G$.
 A  convenient
way of doing this, which at the same time
answers the question posed above, consists
 in considering, instead of bundles, the respective modules
of sections. In this setting, the pull-back
 operation leads
to the following isomorphism of $C(M)$-modules:
 $\Gamma(q^*\xi)\cong C(M)\otimes_{C(M/G)}\Gamma(\xi)$.
Recalling now that $C(M)$ has a decomposition into
$C(M/G)$-submodules we expect, when we regard $\Gamma(q^*\xi)$
as a $C(M/G)$-module, to find a submodule inside it which
is isomorphic
to $\Gamma(\xi)$. This is true, and the submodule we are
looking for can be characterized
in the following way. First notice that the natural $G$-action on $q^*(\xi)$
induces one on $\Gamma(q^*\xi)$. Then we have:
\begin{proposition}(cf.\cite{R})\label{prop:5} The space
of $G$-invariant sections of $\Gamma(q^*\xi)$ is isomorphic,
as a $C(M/G)$-module, to $\Gamma(\xi)$:
\[
\Gamma(\xi)\cong \Gamma^{inv}(q^*\xi)
                  =\{s\in \Gamma(q^*\xi) \;|\; g\cdot s = s
                                 \;\,\mbox{for all}\;\,g\in G\}.
\]
Similarly, if  $\eta$ is a $G$-bundle on $M$ (with action $\tau$),
then the space of $\tau$-invariant sections of $\Gamma(\eta)$
is isomorphic, as a $C(M/G)$-module, to
$\Gamma(\eta/\tau)$.
\end{proposition}
We thus arrive at the conclusion that, in order to
{\it regard}  $\Psi (\in\Gamma(\eta^S))$ as a section on
 a bundle over $\BbR \BbP^2$, we must:
 ({\it i\,}) Specify a $\BbZ_2$-action $\tau$ on
$\eta^S$ and ({\it ii\,})
 Require that $\Psi$ be a $\tau$-invariant section. Regarding
 ({\it i\,}), there are two choices\footnote{On $\BbR\BbP^2$ there
 are only two  equivalence classes of bundles of a given rank $k+1$, represented
 respectively by $L_+^{k+1}$ and $L_-\oplus L_+^k$ (notice that $L_-^2$ is trivial). For the SS problem, the choice
 of connection is also relevant. This is closely related to a choice of
 representative for the class of the bundle. But for each
 class only one choice is compatible with the Fermi-Bose alternative.
 These are the ones we are considering.},
  that can be described as follows. Let $t$ denote the non trivial element of
$\BbZ_2$ and choose an integer $\widetilde K$. Recalling that the total space
of $\eta^S$ is given by
$E(\eta^S)=\{(r,\sum_{J,m_J}\lambda_{J,m_J}|J,m_J(r)\rangle)\;|\;r\in S^2,\,
\lambda_{J,m_J}\in \BbC\}$, we may define an action $\tau:\BbZ_2\times
E(\eta^S)\rightarrow E(\eta^S)$ by setting
\begin{equation}\label{eq:K2}
\tau_t\left(r,\,\lambda|J,m_J(r)\rangle\right):=
\left(-r,(-1)^{2S-J+\widetilde K}\lambda|J,m_J(-r)\rangle\right).
\end{equation}
The induced action on the \emph{section} $\Psi$ gives:
\begin{eqnarray}
(t\cdot \Psi)(r) & := &\tau_t\Psi(-r) =
 \tau_t\Bigl(-r,\,\sum_{J,m_J}\Psi_{j,m}(-r)|j,m(-r)\rangle\Bigr)
\nonumber\\
& = & \Bigl(r,\,\sum_{J,m_J}\Psi_{J,m_J}(-r)
(-1)^{2S-J+\widetilde K}
|J,m_J(r)\rangle\Bigr). \nonumber
\end{eqnarray}
Taking now
({\it ii\,}) into account, we must require $t\cdot \Psi=\Psi$.
This  implies
$\Psi_{J,m_J}(-r)=(-1)^{2S-J+\widetilde K}\Psi_{J,m_J}(r)$.
We are now in a position to discuss the singlevaluedness condition
(\ref{eq:singlevalued}). Let us consider a spin basis
 with the exchange property
$|\overline M(-r)\rangle =(-1)^K |M(r)\rangle$. In the $\{J,m_J\}$ basis
this corresponds to
\begin{equation}\label{eq:Berry-exchange2}
|J,m_J(-r)\rangle=(-1)^{2S-J+K}|J,m_J(r)\rangle.
\end{equation}
We then see that,
with a basis satisfying (\ref{eq:Berry-exchange2}),
 imposing (\ref{eq:singlevalued})  amounts  to require
$\Psi$ to be an invariant section with respect to the action (\ref{eq:K2}),
{\it provided} we choose $K=\widetilde K$.

But note that for the definition of $\tau$ in (\ref{eq:K2}) a
previously specified relation between $|J,m_J(r)\rangle$
and $|J,m_J(-r)\rangle$ is not needed. In fact, not even a
dependence of the basis on $r$ is required, given
that $\eta^S$ is  anyway a trivial bundle.
Hence,
(\ref{eq:Berry-exchange2})
seems not to have a further meaning. Its only role
is to ensure that the fibers of $\eta^S$ at opposite points on the
sphere \emph{coincide}, given that $\eta^S$ was constructed as a
twisted (yet  trivial) subbundle
of a trivial bundle. It then makes sense to ``compare'' the values of the section
$\Psi$ at different points, as is tacitly assumed in  (\ref{eq:singlevalued}).

Due to  proposition \ref{prop:5},
there is an isomorphism
$\Phi:\Gamma(\eta^S/\tau)\rightarrow \Gamma^{inv(\tau)}(\eta^S)$
 of $C(\BbR \BbP^2)$-modules.
 The condition $t\cdot \Psi=\Psi$ (w.r.t $\tau$)
guarantees that $\Psi=\Phi(\sigma)$ for a unique
$\sigma \in\Gamma(\eta^S/\tau)$.
In particular, note that if in (\ref{eq:Berry-exchange2}) we choose
$K=\widetilde K+1$, then that {\it same} section $\sigma$
will be represented on the sphere by a
function $|\Psi'\rangle$ satisfying $|\Psi'(-r)\rangle = - |\Psi'(r)\rangle$
and hence in
contradiction with the singlevaluedness assumption.
%
\subsection{Spin operators}
%
%
In the BR approach, spin operators do also depend on $r$. As
with the spin basis, they are defined making use of the map $U$:
\begin{equation}\label{eq:spin-operators1}
S_i(r):= U(r) S_i U^\dagger(r).
\end{equation}
The spin operators are defined in such a way that they act linearly
-as the physically correct  representation-
on each fiber (recall that in general the fibers are isomorphic to
$V^s\otimes V^s$).
In order to relate this definition to our approach, let us recall
 that, for a given value of $S$, the two bundles corresponding
to Fermi and Bose statistics are $SU(2)$-equivariant. Now, if  under
finite rotations the wave function transforms according to such an
$SU(2)$-action, then the spin operators should correspond to an infinitesimal
version of the corresponding $SU(2)$-action. In the definition of such
infinitesimal operators one must  be careful that only the
{\it spin} degrees
of freedom are being described.
One would perhaps expect that
such a requirement imposes a restriction on the admissible
bundles where the wave function is supposed to be defined.
But  this
is not the case: a consistent description
of identical particles having the physically wrong statistics
is in fact possible within the present approach. This  may be regarded as
a further indication that additional physical requirements
are really needed for a proof of SS in NRQM.

Let us, as an example, consider the case of two spinless particles
obeying fermionic statistics. In that case, as we have seen, the wave
function is defined on $L_-$. Let $\nabla$ be the corresponding  flat
connection and $\tau$  the $SU(2)$-action.
Consider the integral curve
$\gamma_i(t)$ of the projection to $\BbR\BbP^2$ of the vector field $L_i$ on
 the sphere,
with $\gamma_i(0)=\left[x\right]$.
Given  $\left[x\right]\in \BbR \BbP^2$, there is for $t$ small enough,
an element $g_t\in SU(2)$ (unique up to elements in the stability group of
$\left[x\right]$) such that $g_t\cdot\left[x\right]=\gamma_i(t)$.
Given $y$ a vector in the
fiber  over $\left[x\right]$, consider $\tau_{g_t}(y)$.
Parallel-transport this vector
from $g_t\cdot\left[x\right]$ back to $\left[x\right]$ and call
$y_t$ the result.
The local spin operator $S_i(\left[x\right])$ corresponding to a rotation about
the $i^{th}$ axis can then be defined through
\begin{equation}\label{eq:spin-operators2}
S_i(\left[x\right])(y):= \lim_{t\to 0}\frac{i}{t}(y_t -y).
\end{equation}
But because of proposition \ref{prop:3} the operators $S_i(\left[x\right])$
are all equal to zero, as required for scalar particles.

In the case of general $S$, the bundles corresponding  to
Fermi and Bose statistics carry respective flat connections and $SU(2)$
actions. We can therefore take (\ref{eq:spin-operators2}) as
a definition of spin operators. Again because of
proposition  \ref{prop:3}, one sees that no inconsistency arises
in the non-physical case. On the other hand, for the bundle corresponding
to the physically correct SS relation, one gets the same operators
defined in BR by means of (\ref{eq:spin-operators1}).

\section{Discussion}
In this work we have tried to approach the SS problem from a new point of view
which, although formally equivalent to the more familiar ones, enables a clear
formulation and understanding of the problem. This has been illustrated through
a comparison with the BR construction, where our formalism proves to be a much
more natural one (cf. section {\bf 3.2}). Particular attention has been devoted
to the meaning of their singlevaluedness condition, which we have shown to be
misleading. The reason for this is that their construction is actually
performed in the universal cover $\widetilde{\mathcal Q}$  of the configuration
space ${\mathcal  Q}$.
We have shown in a precise way what are the requirements
that allow us to regard sections  defined on bundles over
$\widetilde{\mathcal Q}$ as sections defined on bundles
 over  $\mathcal Q$.
Our approach also settles the question of
how many different constructions of the BR kind exist. Indeed, the
decomposition of  $C(\widetilde{\mathcal  Q})$ into  $C(\mathcal Q)$-submodules
already contains all the necessary information. Moreover, it allows one to work
directly with {\it functions} on $\widetilde{\mathcal Q}$, thus making the
construction of a transported spin basis unnecessary.

\bibliographystyle{plain}

\end{document}